%% file: paper.tex
\PassOptionsToPackage{dvipsnames}{xcolor}
\documentclass[nonacm,format=acmsmall,screen]{acmart}

\usepackage[all]{nowidow}
\usepackage{xcolor}
\usepackage{subcaption}
\usepackage{hyperref}
\usepackage{acronym}
\usepackage{makecell}
\usepackage{nicefrac}
\usepackage{tabularx}
\usepackage{csquotes}
\usepackage{siunitx}
\usepackage{eurosym}
\microtypesetup{expansion=false}
\usepackage[capitalise,noabbrev]{cleveref}
\crefformat{section}{\S#2#1#3}
\crefformat{subsection}{\S#2#1#3}
\crefformat{subsubsection}{\S#2#1#3}
\crefrangeformat{section}{\S#3#1#4 to~\S#5#2#6}
\crefrangeformat{subsection}{\S#3#1#4 to~\S#5#2#6}
\crefrangeformat{subsubsection}{\S#3#1#4 to~\S#5#2#6}

\usepackage{siunitx}
\usepackage{layouts}

\newcommand{\tRange}[3]{\qtyrange[range-units = single, range-phrase = --]{#1}{#2}{#3}}

\begin{document}
\author{Trever Schirmer}
\affiliation{%
    \institution{Technische Universität Berlin}
    \city{Berlin}
    \country{Germany}}
\email{ts@3s.tu-berlin.de}
\orcid{0000-0001-9277-3032}

\author{Aris Wiegand}
\affiliation{%
    \institution{Technische Universität Berlin}
    \city{Berlin}
    \country{Germany}}
\email{aw@3s.tu-berlin.de}
\orcid{0009-0004-5170-3838}

\author{Lucca di Benedetto}
\affiliation{%
    \institution{Technische Universität Berlin}
    \city{Berlin}
    \country{Germany}}
\email{di.benedetto@3s.tu-berlin.de}
\orcid{0009-0004-0807-0676}

\author{Linus Gustafsson}
\affiliation{%
    \institution{Technische Universität Berlin}
    \city{Berlin}
    \country{Germany}}
\email{lig@3s.tu-berlin.de}
\orcid{0009-0001-7557-8742}

\author{Natalie Carl}
\affiliation{%
    \institution{Technische Universität Berlin}
    \city{Berlin}
    \country{Germany}}
\email{nc@3s.tu-berlin.de}
\orcid{0009-0000-5991-9255}

\author{Tobias Pfandzelter}
\affiliation{%
    \institution{Technische Universität Berlin}
    \city{Berlin}
    \country{Germany}}
\email{tp@3s.tu-berlin.de}
\orcid{0000-0002-7868-8613}

\author{David Bermbach}
\affiliation{%
    \institution{Technische Universität Berlin}
    \city{Berlin}
    \country{Germany}}
\email{db@3s.tu-berlin.de}
\orcid{0000-0002-7524-3256}

\title[New Kids]{New Kids: An Architecture and Performance Investigation of Second-Generation Serverless Platforms}



\setcopyright{rightsretained}
\acmJournal{TOIT}
\acmYear{2026}
\acmVolume{1} \acmNumber{1} \acmArticle{} \acmMonth{1} 
\acmDOI{10.1145/3841482}

\begin{abstract}
    With the ever-increasing usage of serverless computing in both industry and academia, it is essential to understand the mechanisms that power the underlying platforms.
    As serverless is more than ten years old, there are different platforms with vastly different approaches.
    We show that, next to the traditional and popular platforms, a second generation of serverless platform has emerged.
    While first-generation platforms are based on containerized, centralized execution, the new generation leverages lightweight isolates and edge deployment.
    This evolution reduces warm request latency from approximately \qty{40}{ms} to around \qty{10}{ms} and reduces cold starts to an afterthought, but limits the execution environment.
    In this paper, we gather and analyze all publicly available information to provide detailed insights into the underlying architecture of seven platforms and then run a microbenchmark-based evaluation totaling more than 38 million function calls to gain a deeper understanding their performance.
\end{abstract}

\makeatletter
\renewcommand{\@authorsaddresses}{%
  \bgroup
  \footnotesize
  © 2026 Copyright held by the owner/author(s).
This is the author’s version of the work. It is posted here for your personal use. Not for redistribution.
The definitive Version of Record was published in ACM Transactions on Internet Technology, \url{https://doi.org/10.1145/3841482}.
  \egroup
}
\makeatother

\maketitle
\renewcommand{\shortauthors}{T.~Schirmer et al.}

\input{sections/1_intro}
\input{sections/2_background}
\input{sections/3_platform_arch}
\input{sections/5_benchmark_arch}

\input{sections/6_benchmark_results}
\input{sections/7_discussion}
\input{sections/9_conclusion}

\begin{acks}
Partially funded by the \grantsponsor{DFG}{Deutsche Forschungsgemeinschaft (DFG, German Research Foundation)}{https://www.dfg.de/en/} -- \grantnum{DFG}{495343202}.
\end{acks}

\bibliographystyle{ACM-Reference-Format}
\bibliography{bibliography.bib}

\end{document}

%% file: sections/1_intro.tex
\section{Introduction}
\label{sec:introduction}

In 2014, Amazon Web Services (AWS) Lambda launched as one of the first commercial serverless platforms, enabling users to simply upload code that should be executed for incoming events.
Since then, the serverless paradigm has massively increased in popularity and is being used in all kinds of use cases to decrease the mental burden on developers, abstract away complex resource mechanisms, and enable highly scalable and elastic workloads.
While it originally started for workloads with infrequent usage~\cite{Wang_2018_Peeking}, its use cases have extended from high performance computing in the datacenter~\cite{Chard_2020_funcX} to the resource-constrained edge~\cite{paper_pfandzelter2020_tinyfaas}.
Many cloud providers have started offering their own serverless platforms, providing their own interpretation of what a serverless platform should look like.

In general terms, serverless platforms execute user code in isolated environments on co-located machines.
The architectural decisions of platforms heavily influence the end user experience:
The isolation mechanism used to separate function instances influences startup overhead and execution duration, as resources shared across many instances leads to constant context switches.
Other aspects, such as the path taken by an incoming request to an instance influences the call overhead, and the process of starting instances influences how long cold starts take.
These happen whenever the platform does not have an already running and idle instance to respond to a request, which often requires downloading the source code from a remote location and starting a new isolated execution environment.
Depending on platform architecture, this can be sped up or even entirely eliminated.

In 2018, Wang et al. published an in-depth investigation on how the cloud providers popular at the time worked and evaluated their resource management and isolation techniques~\cite{Wang_2018_Peeking}.
More than eight years later, the platforms that existed in 2018 have changed considerably, and new platforms have emerged.
These new platforms have a different architecture, use other underlying technologies, and focus on other kinds of use cases, setting different priorities when trading off performance features.
The architecture of commercial platforms is often not well understood, as they are closed-source.
It is however important for researchers and users to understand their differences to decide where to put workloads, and to be able to 
While researchers often use these platforms to build and evaluate their research prototypes, there is no up-to-date research on their inner workings, which is required to gain a better understanding of the evaluation results.

In this paper, we study the underlying architecture of commercial serverless platforms that have existed since the beginning of serverless as well as newcomers that have received less attention in academia so far.
We gather information from published literature, the documentation and various other publications of the cloud providers, as well as our own experiments to get an in-depth understanding of their inner workings.
We then perform microbenchmarks to evaluate the impact of these architectural decisions on end user performance.

Our results show that the first generation of platforms focused on offering event-based general purpose compute with use cases where hundreds of milliseconds of delay are acceptable.
In recent years, a second generation of platforms has emerged that specialize on low latency and fast cold starts.
To achieve these goals, they use new kinds of isolation mechanisms and are geo-distributed.
Additionally, the platforms that have existed for longer have been adapted over the years to add features or improve performance in core areas without re-architecting the whole platform from scratch.
While performing the microbenchmarks, we also noticed that some platforms, most notably Fly.io, had availability and performance issues, demonstrating the need to measure the claims of providers.
We provide our benchmarking code and results as open source\footnote{\url{https://github.com/3s-rg-codes/faas-bench}, \url{https://github.com/3s-rg-codes/faas-bench-explore}}\!\!.

%% file: sections/2_background.tex
\section{Background and Related Work}
\label{sec:background}
Serverless is a computing paradigm in which operational concerns are abstracted away from users~\cite{Manner_2023_Define,Kounev_2023_Definition}.
This includes not needing to decide where and how compute resources are placed or started and having a simple interface for management operations~\cite{jonas2019berkleyServerless}.
Keeping in line with the goal of simplicity for users, billing is often pay-as-you-go.
To reduce their cost, serverless providers usually pool resources between multiple tenants using virtualization~\cite{li2022survey}.
Function as a Service (FaaS) is a serverless offering where users specify code (a \emph{function}) that is to be run in response to incoming events.
When no \emph{function instance} is available for processing an event, a new instance is created in a \emph{cold start}.
These instances are then reused for subsequent requests (\emph{warm starts}) and scaled up and down by the platform, depending on load.
Since many competing definitions for the terms \enquote{serverless} and \enquote{Function-as-a-Service} exist, all aforementioned characteristics can be violated still being considered as serverless or FaaS~\cite{Manner_2023_Define}.
In this paper, we define serverless in the spirit of duck typing: If a cloud service provider considers their product to be serverless, it probably is.
Moreover, we use the terms \enquote{serverless} to indicate a focus on abstracting from operational concerns and \enquote{FaaS} when the focus lies in code execution.

Lin et al. split FaaS platform architectures into the following layers~\cite{li2022survey}.

\paragraph{Virtualization}
Serverless platforms isolate functions by relying on virtualization technologies.
Traditional virtual machines are ill-suited for the isolation between functions due to excessive start-up times~\cite{Agache_2020_Firecracker}.
Instead, \emph{microVMs} --- stripped-down virtual machines that only support the bare minimum of features --- are used for function instances, reducing cold starts to \textasciitilde\qty{100}{ms}~\cite{Brooker_2021_Restoring,Agache_2020_Firecracker}.
Containers lie one layer above microVMs and present a more lightweight, kernel-based isolation mechanism, where isolates share the underlying operating system and infrastructure.
While Docker remains the most popular container engine, serverless platforms --- such as Google Cloud Run Functions --- commonly use gVisor, an OCI-compatible container runtime that trades performance for stronger isolation by bringing system calls into the user space~\cite{Debab_Hidouci_2021,website_gvisor,oci2026oci}.
On an even higher level, it is also possible to use the programming language runtime for isolation.
For example, instead of executing JavaScript with a traditional runtime, such as Node.js, code can be executed in the same isolates used by the Chrome browser's V8 engine for separating browser tabs.
In V8, every isolate has its own global variables, heap, and state, but context-switches happen within one process~\cite{website_v8Isolates}.
Another possibility is running code compiled to WebAssembly, which requires even less isolation because it is always side effect free~\cite{website_wasmtime}.
Generally, the higher up the isolation lies, the lower its overhead --- while also adding constrains that users have to consider.

\paragraph{Encapsulation}
This layer manages the virtualization layers of the platform's worker nodes.
It starts and stops instances and routes requests to them.
Optimizations on this layer include deciding when instances need to be started, e.g., by pooling function instances, predicatively starting instances, and caching instances and by optimizing how the function code is loaded to the worker during instance start-up~\cite{Brooker_2023_OnDemand,Du_2020_CatalyzerSubMillisecond}.

\paragraph{Orchestration}
The orchestration layer is responsible for routing requests to appropriate worker nodes.
Centralized orchestrators optimize placement through global knowledge but are difficult to scale and introduce a single point of failure~\cite{Cvetković_2024_Dirigent}.
Distributed approaches have less optimal placement and introduce communication overheads but are more scalable~\cite{paper_pfandzelter2020_tinyfaas}.
Depending on the platform, the orchestration layer can take over additional responsibilities, such as keeping instances alive to minimize cold starts as well as load balancing between them.

\paragraph{Control plane}
The control plane handles function data as well as metadata, such as resource limits.
Users interact with the control plane when creating or updating functions.
The data then needs to be shared with the orchestration layer and workers:
either by pushing to nodes, leading to high overhead for functions with low usage, or by workers pulling when necessary, leading to slow cold starts~\cite{Brooker_2023_OnDemand,Cvetković_2024_Dirigent}.

\subsection{Related Work}
While this paper focuses on the architecture of \emph{commercial} serverless platforms, there exist many research platforms~\cite{paper_pfandzelter2020_tinyfaas,Chen_2024_YuanRong,Chard_2020_funcX,schirmer2025hyperfaas}.
These often focus on improving a specific aspect of serverless platforms
~\cite{Wen_2023_Rise}. 
Other research focuses on understanding open source platforms, which are often based on Kubernetes~\cite{Li_2019_OpenSourcePlatforms}.
Commercial serverless platforms have previously been analyzed extensively, e.g., studying their performance over different durations~\cite{Eismann_2022_LongTerm,schirmer2023nightshift} and focusing on comparing different aspects of the platforms, such as their billing model~\cite{Lin_2025_Billing}.
However, current research focuses almost exclusively on the most well-known and popular platforms, not the newer and platforms.
To better understand the direction serverless platforms are moving towards, \citet{Wen_2023_Rise} have conducted a systematic literature review of 164 papers, identifying 17 possible directions for future research.

\section{Methodology}
The choice of platforms to study in-depth was made based on multiple factors.
First, we included two of the well-established and popular platforms studied in 2018 by Wang et al.~\cite{Wang_2018_Peeking}: AWS Lambda and Google Cloud Run Functions.
They also studied Azure Functions, which we exclude in this paper since it has consistently underperformed in previous research~\cite{Ustiugov_2021_Stellar,Wang_2018_Peeking} and thus has less relevance for understanding how platforms might evolve.
We then asked industry experts and serverless researchers for a list of serverless platforms they know and searched the internet for additional platforms.
We narrowed down this list by removing platforms that build on top of other serverless platforms, as the underlying architecture is the same.
For example, multiple serverless platforms are built on top of AWS Lambda~\cite{website_netlifyIsAWS,website_vercelIsAWS}.
Afterwards, we analyzed the remaining platforms for whether they have enough publicly accessible information to infer their underlying architecture and removed platforms that did not offer enough results, as in the case of Alibaba Function Compute and Huawei Cloud Functions.
To determine whether a platform has enough available documentation, we searched for publications from the particular company regarding their platforms and also searched the internet for other sources of information, e.g., documentation, the company blog, and conference talks.
We additionally filtered out platforms that use Kubernetes~(K8s), a popular cluster scheduler~\cite{Rensin_2015_KubernetesWhitepaper}, with the default scheduling mechanism.
Research into the scheduling mechanism of Kubernetes has demonstrated that it underperforms for serverless workloads, as it was designed for long-running workloads with higher startup delays~\cite{Cvetković_2024_Dirigent}.
\cref{tab:platformsIncluded} shows an overview of all platforms that were included in the in-depth study.

{
    \newcommand{\VM}{µVM}
    \begin{table*}[]
        \begin{tabularx}{\linewidth}{>{\raggedright\arraybackslash}X c l l c}
            \toprule
            \bf{Platform} & \bf{Year} & \bf{Isolation} & \bf{Locality} & \bf{Ref.} \\
            \midrule
            AWS Lambda                                  & 2014 & \VM{}           & regional  & \cite{aws2014lambda}      \\
            Google Cloud Run Functions (CRF)$^\beta$    & 2017 & container/\VM{} & regional  & \cite{google2017gcf}      \\
            Oracle Cloud Infrastructure Functions (OCF) & 2019 & container/\VM{} & regional  & \cite{oracle2019ocf}      \\
            \midrule
            Cloudflare Workers (CW)                     & 2020 & intra-process   & worldwide & \cite{cf2020workers}      \\
            Deno Deploy (DD)                            & 2021 & intra-process   & worldwide & \cite{denoland2021deploy} \\
            Fastly Edge Compute (FEC)                   & 2019 & intra-process   & worldwide & \cite{fastly2019edge}     \\
            Fly.io$^\alpha$                             & 2020 & \VM{}           & worldwide & \cite{flyio2020fly}       \\
            \bottomrule
        \end{tabularx}
        \caption{
            The platforms included in this paper were chosen based on the previous work by Wang et al.~\cite{Wang_2018_Peeking} and by researching accessible commercial serverless platforms with sufficient documentation.
            Platforms with regional locality have different regions, of which users choose one.
            Worldwide platforms create replicas and choose the closest for answering client requests.
            CW, DD, FEC, and Fly.io have \textasciitilde{}300, 6, \textasciitilde{}80, and 35 replica locations, respectively.
            \footnotesize{
                $^\alpha$Regional configuration is also possible.
                $^\beta$A second version was released in 2024.
            }
        }
        \label{tab:platformsIncluded}
    \end{table*}
}

%% file: sections/3_platform_arch.tex
\section{Platform Architectures}
\label{sec:arch}
For the platforms we selected to study in-depth, we use the collected information to provide an overview of their architecture.
We begin by characterizing the regional platforms and then move on to the platforms with worldwide deployments.

\subsection{AWS Lambda}\label{arch:lambda}

\begin{figure*} 
    \centering
    \includegraphics{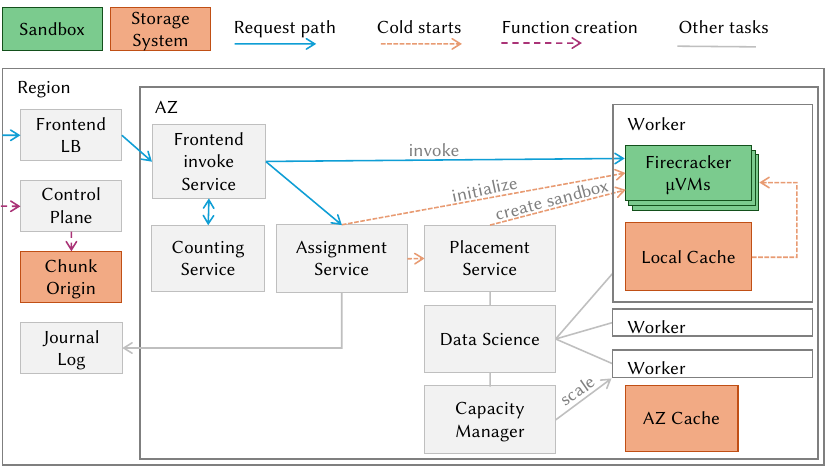}
    \caption{
      AWS Lambda architecture and request workflow.
      Within a region, requests are load-balanced to an availability zone.
      Within an availability zone, requests are assigned to instances by multiple services running inside that zone, and a per-region control plane.
      Function data is stored in a per-region store and cached in the availability zone as well as the workers.
    }
    \Description{Architecture Overview. Graph with Lines showing control and request flow.}
    \label{fig:arch:lambdaRequest}
\end{figure*}

\emph{AWS Lambda}, introduced in 2014, is widely considered to be the first and most well-studied FaaS platform~\cite{Leitner_2019_MixedMethod}.
\Cref{fig:arch:lambdaRequest} shows an overview of its architecture and request workflow.

\subsubsection{Resource Model}
Functions are created by talking to the control plane of a region and either uploading the source code or pointing to an OCI image that can be up to 10 GiB large.
Internally, the image is deterministically turned into an ext4 file system, which is then split into chunks and uploaded to the internal storage system~\cite{Brooker_2023_OnDemand}.

\paragraph{Sandboxing}
AWS has open-sourced the Virtual Machine Manager~(VMM) it uses in Lambda to manage microVMs, called \emph{Firecracker}~\cite{Agache_2020_Firecracker}.
The VMM itself runs as an unprivileged process on the host OS and is limited to a minimal set of syscalls.
As it is a Kernel-based VM~(KVM), it can only be used to run Linux binaries on Linux hosts using hardware-assisted system-level virtualization~\cite{Agache_2020_Firecracker}.
Lambda further hardens this boundary by wrapping Firecracker in a \emph{jailer} that enforces cgroups, namespaces, and seccomp filters on the host, minimizing the Trusted Computing Base~\cite{Agache_2020_Firecracker}.
Despite this, shared silicon remains susceptible to microarchitectural side-channel attacks, necessitating mitigations such as disabling Simultaneous Multithreading~(SMT).

\paragraph{Language Runtimes}

Lambda provides managed runtimes for different languages~\cite{website_awsRuntimes}, which manage interaction between the platform and function code, e.g., deserializing the request object into the native representation.
Uploading an OCI image allows using any other programming language, but requires functions to implement the Lambda runtime API themselves.

\paragraph{Locality}

Lambda is a regional service, with a function created in a specific region~\cite{website_awsRegions}.
Inside a region, requests to a function are balanced between multiple availability zones.
To achieve this, the URL that is used to call a function includes the region name.

\paragraph{Configurability}

Users can configure the amount of memory a function has access to between \qty{128}{MB} and \qty{10240}{MB}, with the platform proportionally scaling compute and other resources~\cite{website_awsMemory}.
To get access to \qty{1}{vCPU}, \qty{1769}{MB} of memory need to be configured.
Users can choose between x86 or ARM CPUs, with ARM-based CPUs offering a lower base price.
The instance concurrency is always set to one, i.e., a function instance only handles one parallel request.
The timeout of a function can be configured between three seconds and 15 minutes.

\paragraph{Pricing}

Lambda bills per request as well as per megabyte-millisecond of runtime.
It is possible to have background jobs running inside function instances, in which case the function instance is billed for the whole time the job is running.

\subsubsection{Request Workflow}

Every region comprises multiple availability zones~(AZ), with requests being load-balanced between zones by \emph{Load Balancers} that also handle TLS termination.
Requests are forwarded to a per-AZ \emph{Frontend Invoke Service}, a stateless service that handles authentication and authorization as well as loading metadata associated to the request.
The Frontend Invoke Service then communicates with the \emph{Counting Service}, a stateful service that enforces global quota limits.
The \emph{Assignment Service} manages the state of a fleet of sandboxes located on different worker hosts.
It determines the sandbox located on a specific \emph{Worker} to execute the function in, so that the Frontend Invoke Service can route the request to it.
If there is no available sandbox, it communicates with the \emph{Placement Service} to initiate a scale-out~\cite{website_awsLambdaUnderTheHood,website_awsCloserLook}.
The Assignment Service is split into partitions, each comprising one Leader and two Followers placed in different AZs to prevent sandbox state being lost in case of failures.
In order to replicate the state of sandboxes managed by one partition across multiple AZs, the Leader of the partition writes to an external \emph{Journal Log Service} that is read by the Followers through a log stream.
The number of partitions and instances of the Assignment Service can be scaled up in response to load spikes by the control plane, which manages the lifecycle of Assignment Service instances~\cite{website_awsLambdaUnderTheHood,website_awsCloserLook}.
The worker hosts are bare metal Amazon \emph{EC2 AWS Nitro}~\cite{aws2026nitro} instances that can host up to 4000 sandboxes~\cite{website_awsNetworkingDetails}.

\subsubsection{Cold Starts}
\begin{figure}
    \centering
    \includegraphics{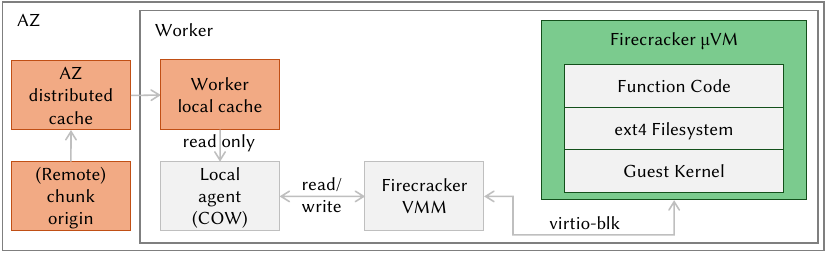}
    \caption{
      On-Demand data loading implementation in AWS Lambda that allows lazily downloading only the chunks of function data that is actually used~\cite{Brooker_2023_OnDemand}.
      The system works by creating a local copy-on-write file system that stores the function data.
      This file system loads missing chunks on-demand when they are actually loaded by the function.
    }
    \Description{Graph showing the tiered caches that are used in data loading. The origin chunk loads to an AZ-Cache, which loads into a Woker-Local Cache. Inside the Worker, a COW-file system is created that is mounted by the Firecracker VMM into the guest kernel.}
    \label{fig:arch:lambdaCaching}
\end{figure}

To start up a function instance in AWS Lambda, a Firecracker microVM has to be booted, function code and dependencies have to be downloaded, and the language-specific runtime has to be initialized before the handler method can be invoked~\cite{Brooker_2021_Restoring}.
To minimize the boot time, workers keep a pool of pre-booted VMs without any function code.

\paragraph*{On-demand loading}
To reduce the download time, Lambda implements a storage and caching system that allows for data to be fetched on-demand~\cite{Brooker_2023_OnDemand}.
Before its introduction, functions were limited to a size of 250 MB, which in case of a cold start was fully downloaded from S3 before execution could start.
With the new system shown in~\cref{fig:arch:lambdaCaching}, data is fetched as needed, so that only the minimal required chunks are downloaded before execution.
MicroVMs for a function have access to two devices emulated by the Firecracker VMM: A root device is read-only and shared between all instances, and one device is passed to a local agent that implements a copy-on-write~(COW) file system that stores instance-specific data.
If a data chunk is not in the local cache, the local agent will first try downloading it from the AZ-level cache, then from the origin server and add it to the local or AZ-level cache if it was not present.
Due to using a COW file system, the chunks can be re-used.
This design leads to two kinds of cold starts: The first time a function is invoked in an AZ or on a specific worker, the caches need to be prepopulated.
Subsequent cold starts on the same worker might be faster, as the files can be read directly from the worker local cache.
These caches are shared between users, which leads to faster startup times when using runtimes provided by the platform since their chunks are likely to be cached already.

\paragraph*{Snapshots}
Java is known for having performance issues in serverless environments due to its long startup times, with initialization taking up between \qty{25}{\percent} and \qty{95}{\percent} of startup time~\cite{Brooker_2021_Restoring,Du_2020_CatalyzerSubMillisecond, Zhang_2019_StorageFunctions}.
To reduce startup times in these cases, Lambda supports \emph{SnapStart} for Java functions:
During function creation, the function is fully booted, and then a snapshot of memory is taken and distributed in the same way that the raw source is distributed.

\paragraph*{Load prediction}
Lambda uses reinforcement learning and deep learning to scale worker hosts and function instances based on the predicted number of incoming requests~\cite{Brooker_2021_Restoring}.
They claim that this leads to less than one percent of invocations being cold starts in production workloads~\cite{website_awsExecutionEnvironment}.
They also use machine learning to optimize packing density of workers, aiming to schedule diverse functions in one worker as they are less likely to contend for the same resources at the same time~\cite{website_awsCloserLook,website_awsLambdaUnderTheHood}.

\paragraph*{Pre-warmed instances}
Lambda can keep a minimum number of warm instances to reduce the impact of a scale from zero.
Users are billed for the whole time the instances are running.

\subsection{Google Cloud Run Functions}\label{arch:gcf}

\begin{figure*}
    \centering
    \includegraphics[width=1.0\textwidth]{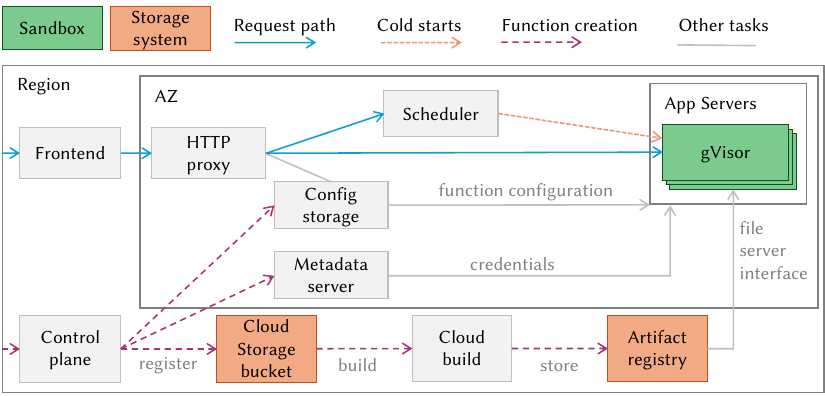}
    \caption{Cloud Run Architecture.
    Requests and management calls are terminated at a per-region front end service, which load-balances to multiple availability zones.
    In an availability zone, a proxy forwards requests to a function instance supported by a scheduler and configuration store.
    Function data is stored in a per-region artifact registry, which are loaded into the function instance.
    }
    \Description{Architecture Overview. Graph with Lines showing control and request flow.}
    \label{fig:arch:googleRequest}
\end{figure*}

\emph{Cloud Run Functions}~(CRF), previously called \emph{Google Cloud Functions 2nd Gen}, is an abstraction on top of \emph{Google Cloud Run}, Google's managed Container-as-a-Service offering.
Calls to the CRF API are translated to Cloud Run API calls with default values.
It is possible to configure the same function using both APIs, depending on the level of detail that should be configured.
A high-level overview of the platform request workflow is shown in~\cref{fig:arch:googleRequest}.

\subsubsection{Resource Model}

To create a function, users can provide a URL pointing to an OCI image or their own source code.
When uploading source code, the raw code is stored in a bucket in Google's managed object storage \emph{Cloud Storage}.
Another service called \emph{Cloud Build} is then used to build a container image and upload it to the image registry service \emph{Artifact Registry}.
In both cases, the image is then deployed in \emph{Cloud Run} as a \emph{Cloud Run Service}.
The configuration itself is stored using Spanner~\cite{Corbett_2013_Spanner} and Bigtable~\cite{Chang_2008_Bigtable}, two storage systems developed by Google.
Cloud Run itself is API-compatible to Knative~\cite{website_cloudRunKnative, website_gcpCompatibleKNative} but runs on top of Borg~\cite{Verma_2015_GoogleBorg}, the Google-internal predecessor of Kubernetes.
Therefore, some options of Knative that map to Kubernetes primitives not used in Borg are not available in Cloud Run~\cite{website_cloudRunKnative}.
\paragraph{Language Runtimes}
CRF has preconfigured runtimes where incoming requests are automatically transformed to language-specific structures.
Users can also provide an OCI image that exposes a compatible HTTP server.

\paragraph{Locality}
CRF is a regional service, with a function deployed to exactly one cloud region~\cite{website_gcpRunLocations}.
The URL that a function can be called with contains the region name to route requests.
Within one region, requests are automatically load-balanced between multiple availability zones~\cite{website_gcpFunctionsLocations}.

\paragraph{Configurability}

CRF offers users to provision functions with different amounts of memory and scales CPU accordingly.
It is possible to configure the available compute and memory allocation between \qty{128}{MiB}--\qty{32}{GiB} RAM and \tRange{0.08}{8}{vCPUs}, with certain vCPU allocations requiring a minimum memory allocation.
The minimum required memory on Cloud Run depends on the execution environment, meaning whether the instance runs in a sandbox based on gVisor or a microVM~\cite{website_gcpRunExecutionEnvironment}. 
The default timeout of \qty{60}{\second} can be extended to up to one hour, and the instance concurrency can be configured between \qty{1}{} (the default) and \qty{1000}{}.

\paragraph{Pricing}

Cloud Run offers different pricing options.
When deploying a normal serverless application, services are billed per vCPU-second, per GiB-second, and per request.
It is also possible to pay for the whole time the instance is running (e.g., to do background jobs as well), in which case there is no per-request cost and the cost per second for compute and memory is reduced. 

\subsubsection{Request Workflow}

When a request enters a cloud region, it is TLS-terminated and load balanced by the \emph{Front End Service} that is used by all internet-accessible cloud services~\cite{website_gcpFrontendService,website_gcpBackstageVideo}.
In a region, an HTTP proxy forwards requests to function instances hosted in so-called \emph{App Servers}.
This is managed by a scheduler component that assigns function instances to incoming requests and creates a new one if necessary.
Function instances are sandboxed using gVisor or microVMs~\cite{website_gvisor,website_gcpRunExecutionEnvironment}.
When configuring an instance concurrency over one, the platform targets a CPU utilization of \qty{60}{\percent}~\cite{website_gcpInstanceAutoscaling}, i.e., the allowed instance concurrency might never be reached.
The CPU utilization is measured every five seconds, limiting elasticity~\cite{website_gcpInstanceAutoscaling}.

\subsubsection{Cold Starts}
To start a new function instance, the container image is fetched from the Artifact Registry service.
Afterwards, the image is launched by the host using gVisor.
The function instance then needs to complete its own initialization before it can accept requests.
To reduce the duration of cold starts, Container Image Streaming and CPU-Boosts are used.
The frequency of cold starts can be reduced using pre-warmed instances and instance concurrency.

\paragraph{Container Image Streaming}
While the details are not well documented, Cloud Run is streaming the container images to the App Server instead of waiting for the full transfer of the image before starting the container~\cite{website_gcpStartContainersQuickly}, which is also used in the hosted Kubernetes service \emph{Google Kubernetes Engine}~\cite{website_gkeContainerImageStreaming}.
This allows containers to start up before the whole image is transferred, making startup times independent of the total image size.
Images are cached using multiple layers on the network, on disk, and in memory.
Using Image Streaming, the whole image is downloaded to the worker node eventually (this differs from the approach used in AWS Lambda, where data is loaded lazily).

\paragraph{CPU-Boosting}
During start-up and for the first ten seconds of runtime, a function instance is allocated a higher number of vCPUs by default.
Functions have access to at least twice the vCPUs, up to a maximum of \qty{8}{vCPU}.
In our experience, this can lead to cold starts being faster than normal invocations.
Users pay for the allocated vCPUs, so that this does not necessarily reduce cost.
Additionally, the function code does not know its current vCPU allocation, complicating manual cost calculations and performance comparisons.

\paragraph{Pre-Warmed Instances}
To limit the impact of a scale-from-zero, CRF keeps unused instances warm for 15 minutes.
Additionally, users can configure a minimum number of instances that should always be running, for which they pay a reduced execution while they are idle.

\paragraph{Instance Concurrency}

Within a function instance, up to 1000 requests can be handled in parallel.
While the instance concurrency is limited to one by default for better understandability, increasing it can reduce the number of cold starts.
This requires users to have a good understanding of their workload, as it might lead to under- or overutilized instances, or overload the resources of an instance.

\subsubsection{Sandboxing}
CRF uses gVisor~\cite{website_gvisor} to run containers, providing lightweight isolation.
It intercepts syscalls to the kernel and executes them in a user space process, where common syscalls are implemented.
Only few unproblematic syscalls are passed through to the kernel.
This effectively mitigates container escape vulnerabilities at the cost of measurable I/O latency from the additional syscall redirection layer~\cite{young_true_2019}.
CRF has added initial support for using microVMs as a sandboxing mechanism, which enables faster CPU and network performance but has slower cold start times~\cite{website_gcpRunExecutionEnvironment}.
They achieve this by employing a secure, two-layer sandboxing environment using a hardware-backed layer for x86 virtualization with a software kernel layer—such as Google's gVisor user-space kernel—to safely execute and isolate workloads \cite{gcf-code-execution}.

\subsection{OCI Functions}
\label{arch:ocf}

\emph{Oracle Cloud Infrastructure Functions}~(\emph{OCF}) is based on the open-source platform \emph{Fn Project}.
It is the only platform in this paper that is built on top of K8s.
Rather than using the default scheduler that is designed for long-running workloads and thus unsuitable for scheduling short-lived event-based functions~\cite{Cvetković_ManagementCost_2023}, OCF uses K8s to start long-running worker nodes that in turn start containers with the actual function instances.
OCF is the only open source-based platform, making it easy to adapt for research use cases~\cite{website_oracleFunctions}.

\subsubsection{Resource Model}

Functions are created in one of their cloud regions, which stores the OCI image and configuration of the function.
Every function is assigned to a subnet and can access resources inside that subnet.
Functions have an instance parallelism of one and can be called synchronously or asynchronously.
OCF is the only platform in this paper that optionally supports multi architecture functions: \emph{the same} function can be uploaded for x86 and ARM.
In this case, requests are routed based on resource availability.

\paragraph{Sandboxing}
OCF uses Docker for sandboxing.
The function process inside the container does not have root access to the container and has to execute as a pre-specified user and group.

\paragraph{Language Runtimes}

As OCF runs function instances in Docker containers using HTTP as transport protocol, any language can be used.
The platform provides SDKs for common languages to simplify this process.

\paragraph{Locality}
Similar to Lambda and CRF, OCF is regional with functions deployed to one region and the function URL containing this region for routing purposes.
Unlike those two, all regions have the same price.

\paragraph{Configurability}

Users can configure the memory of function instances in steps between \SI{128}{MB} and \SI{3072}{MB}, and the timeout to a maximum of \SI{300}{s}.
While the vCPU assigned to an instance is not documented, we measured our experiment functions and document the results in~\cref{tab:experimentOverview}.

\paragraph{Pricing}

Execution is billed per request and per gigabyte memory-second.
Provisioned capacity is billed at \nicefrac{1}{4} of execution cost when unused.

\subsubsection{Request Workflow}

Requests enter a load balancer, which forwards them to workers using consistent hashing to always send request for a function to the same worker.
If the worker is overloaded and rejects the call, the next worker is tried.
With this architecture, one single worker has to receive every single call to a function (even if just to reject it), limiting scalability~\cite{website_fnGithubPlacement}.

\subsubsection{Cold Starts}
As the load balancer does not track warm function instances on workers, the workers start instances on demand when no free instance is available for a request.
There are no special methods used to decrease the duration of a cold start, which comprises downloading the image from a registry and starting it, after which the worker tries sending the request until the HTTP server inside the container answers.
The platform supports provisioned concurrency to ensure that a specific number of function instances are always available, which are billed at a reduced rate when idle.

\subsection{Cloudflare Workers}\label{arch:cw}
\begin{figure*}
    \centering
    \includegraphics[width=\textwidth]{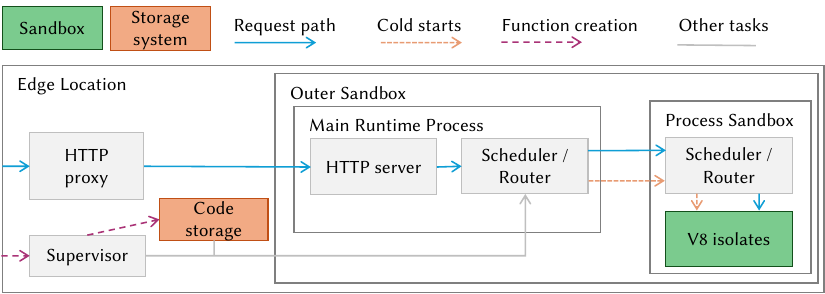}
    \caption{Cloudflare Workers Architecture.
    Requests are terminated at every edge location by an inbound HTTP proxy.
    They are then forwarded to a worker node, which run multiple instances of the V8 runtime.
    Inside a process, an HTTP server receives requests and routes them to a V8 isolate, which is a thread running in its process namespace.
    Function data storage is replicated to every edge location.
    }
    \Description{Architecture Overview. Graph with Lines showing control and request flow.}
    \label{fig:arch:workersRequest}
\end{figure*}

The first platform we analyze with worldwide deployment, Cloudflare offers the FaaS Platform \emph{Cloudflare Workers} (CW) that allows executing code inside a V8 isolate using its globally distributed edge locations.
An overview of its architecture is shown in~\cref{fig:arch:workersRequest}.

\subsubsection{Resource Model}

CW refers to functions as \emph{Workers}, a name based on the concept of \emph{Service Workers} from the JavaScript ecosystem that enables running background tasks client-side.
To create a function, users upload their code, dependencies, and additional files to be compressed into a module with a size of up to \SI{10}{MB}.
This module is then stored on-disk in every edge location within three seconds.
For isolation, it uses V8, leading to native support for JavaScript, WebAssembly and common Web APIs.
Support for additional languages is possible by compiling them to Wasm.
Due to using V8, the initial calling code always needs to be JavaScript or TypeScript, and all Wasm code needs to be side effect free.
Cloudflare has worked on extending the support for other languages such as Python and Rust by providing integrations into the JavaScript APIs.
Resources are not configurable, with a memory limit of \SI{128}{MB} per isolate~\cite{website_cloudflareMem}.
Since the instance concurrency is not configurable, users can not control the resource consumption per call.
If an isolate exceeds its resource usage, it is softly evicted, i.e., it will be removed after all running requests are completed.

\paragraph{Locality}

Functions are deployed to all edge locations, and are usually executed in the edge location the request arrives in.
To mitigate high load, requests can be forwarded to other edge locations with lower load.
The enterprise tier allows limiting the locations the function is deployed to.
The platform analyzes call patterns of workers and places it in an edge location that minimizes the request duration of remote HTTP requests the function itself makes~\cite{website_cloudflareSmartPlacement}.

\paragraph{Sandboxing}

CW uses V8 isolates for sandboxing.
Every server runs multiple runtimes in their own processes, each comprising multiple threads with different isolates.
Most functionalities with side effects are offered using Web APIs that are handled by the runtime, i.e., shared between multiple function instances.
This means that every function instance has its own thread, whose resources are then limited and controlled by the platform.
Additionally, Linux security features such as namespaces and seccomp are used to limit the capabilities of the thread.
Due to their sandboxing mechanism, the security model also differs compared to other platforms.
For example, the JavaScript Date API always returns the date of the last I/O operation to make timing-attacks impossible~\cite{website_cloudflareSecurityModel}, and threading within a function is disabled, both of which complicate benchmarking experiments.
Tenants sharing the same process nevertheless face elevated risk from intra-process side-channel attacks~\cite{marin_serverless_2022}; timer degradation is the primary mitigation, inherently restricting some developer capabilities.
An advantage of this approach is that the marginal cost of an instance that is waiting for another resource (e.g., a Web Request) is very low, as it uses almost no resources except for its own RAM, which is often only a couple of MB, essentially only the size of the code files~\cite{website_cloudflareMem}.

\paragraph{Cost}
Workers are charged per request from the outside and for CPU time.
They do not charge for calls between functions and only charge for \emph{active} CPU time, i.e., the time a CPU is not waiting.
This means that waiting for another request does not incur any cost as the CPU is in a waiting state, eliminating the double-billing problem~\cite{Baldini_2017_Trilemma}.

\subsubsection{Request Workflow}

Using Anycast IP Addresses, requests are routed to an \emph{Inbound HTTP proxy} in the closest edge location, where TLS is terminated.
The request is then forwarded on the local server to a \emph{Scheduler}, which assigns it to a runtime.
The response is then sent through an \emph{Outbound HTTP proxy}.
Calls between two functions can circumvent all of this by using \emph{Service bindings}, which directly call JavaScript methods in other functions.
As the other worker is likely running in the same process or even the same thread on the host, service bindings minimize call overhead.

\subsubsection{Cold Starts}
Cloudflare limits the number of cold starts by keeping function instances alive until they need to be evicted, in which case the least recently used function is stopped.
Additionally, a function instance can already be started by the inbound HTTP proxy before the TLS handshake is finished, as clients send the hostname during the first part of the handshake.
This hostname contains the function name, so that an instance can already be created before any payload is sent (after two roundtrips in the TLS handshake).
When starting a new function, the platform needs to load the source code from disk, start a V8 isolate, and initialize the global scope.
If all of this takes less time than the TLS handshake, the client-side cold start latency is effectively zero.

\subsection{Deno Deploy}\label{arch:deno}
In 2021, Deno released a globally distributed, V8-based FaaS platform called \emph{Deno Deploy} (DD) inspired by and similar to Cloudflare Workers~\cite{website_denoInspired}.
The underlying runtime is open-source~\cite{denoland2026gh}.

\subsubsection{Resource Model}
To create a function, users can upload up to \SI{1}{GB} of files that are then pre-processed by the control plane to minimize the duration of cold starts.
If necessary, the code is transpiled to JavaScript, and all dependencies are downloaded and optimized~\cite{website_denoCloudPlatform}.
The resulting artifact is uploaded to object storage in every cloud region for fast startup times.
The control plane then updates a centralized Postgres database with function information, and updates a proxy running in every cloud region to route requests to the function.
Every function has access to \SI{512}{MB} RAM and is limited to \SI{50}{ms} CPU time per request.
Again, only active CPU time is counted.
The V8 runtime can be used to execute Wasm, so that languages that can be compiled to Wasm are supported.

\paragraph{Sandboxing}
Similarly to CW, DD uses V8 isolates for tenant isolation within a shared host process, minimizing cold start latency and per-instance memory overhead.
Unlike CW, the \texttt{Date} API is supported, enabling timing attacks~\cite{marin_serverless_2022}.

\paragraph{Locality}
DD runs functions in virtual machines running in \num{6} Google Cloud Platform regions~\cite{website_denoRegions}, and a request will always be executed in the closest region using Anycast IPs.

\paragraph{Pricing}

Unlike the other platforms, DD has a fixed price per request, with a fixed maximum active CPU time of \SI{50}{ms}.
Again, asynchronous I/O does not count towards the limit.
Network usage is billed based on actual usage. 

\subsubsection{Request Workflow}
Functions can only be triggered using HTTP calls.
After entering a region, the proxy in this region decides how many runners in this region should be executing the function of this request in parallel.
It then sends the request to the runner with the lowest resource utilization, i.e., that is expected to finish the requests the fastest.
In case of a cold start, the runner downloads the optimized function code from object storage within the same region and starts a new isolate.
The runners only start one isolate per function, leading to potentially high instance parallelism that can not be configured.
The timeout for stopping unused instances is not documented.

\subsection{Fastly Edge Compute}\label{arch:fastly}

Fastly is an internet provider with a global network whose main product is a Content Delivery Network~(CDN).
They offer a FaaS platform called \emph{Edge Compute} (FEC), that launched in 2019, where users can run~WebAssembly code.
Their main use case is transforming requests before they are forwarded to a backend server, and reconfiguring their CDN per request~\cite{website_fastlyDocumentation}.

\subsubsection{Resource Model}

Functions are created by uploading a Wasm application with a size of up to \SI{100}{MB} to the control plane~\cite{website_fastlyDocumentation}.
While FEC supports some programming languages explicitly, it is possible to write the functions in any language that can be compiled to Wasm as long as the program uses the Fastly ABI~\cite{website_fastlyDocumentation}.
Every request has a non-configurable memory limit of \SI{128}{MB}.
Functions can be called via HTTP, with a configurable domain name.

\paragraph{Sandboxing}

FEC uses the Wasmtime~\cite{website_wasmtime} runtime.
Isolation is mostly offered by the language itself: since WebAssembly modules can not have side effects, it is not necessary to use extra isolation mechanisms.
To have side effects, e.g., to interact with outside services, the FEC offers an Application Binary Interface~(ABI) with methods for sending HTTP requests and interacting with other Fastly services.
The API is however limited, e.g., there is no method of getting the current time or interacting with the file system~\cite{website_fastlyDocumentation}.
The absence of a system-time API and WebAssembly's memory-safe execution model inherently constrain the attack surface, providing strong isolation without additional security middleware.

\paragraph{Locality}

FEC uses the edge network of Fastly, with requests being routed to the closest edge location using GeoDNS.

\paragraph{Pricing}

Function costs are calculated with a fixed cost per request, per millisecond of duration, with requests over \qty{20}{\milli\second} being more expensive.

\subsection{Fly.io}\label{arch:fly}

\begin{figure*}
    \centering
    \includegraphics{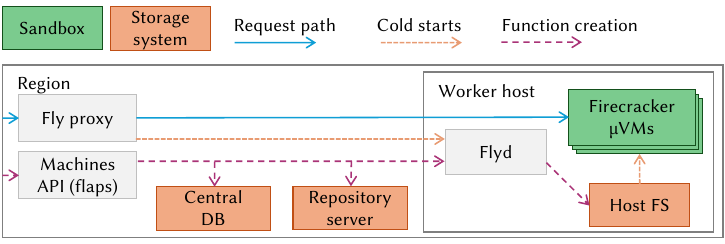}
    \caption{
      Fly.io Architecture.
      In every region, a proxy terminates requests.
      This proxy searches for an idle or underutilized machine and forwards the request to its worker host.
      The worker host starts the machine if necessary.
      Function data is stored in a per-worker cache, which is pulled from a centralized database if necessary.
      }
    \Description{Architecture Overview. Graph with Lines showing control and request flow.}
    \label{fig:arch:flyRequest}
\end{figure*}

Fly.io is a cloud offering created in 2017 and launched in 2020.
Their main offering is running OCI images at edge locations and transparently forwarding traffic to these instances.

\subsubsection{Resource Model}
The main resources are \emph{Apps} and \emph{Machines}.
An App comprises multiple Machines, which are Firecracker microVMs all running an identical image \cite{mackey_fly_machines_2022}.
To create a function, users need to build an OCI image, which is then uploaded to the region where the function is created.
All images need to expose an HTTP server that requests are forwarded to, so there are no specially supported programming languages.
They provide a client-side CLI tool to generate a Dockerfile based on the codebase for common programming languages and frameworks~\cite{flyio2020scanners}.
Users then need to create Machines, which are possible function instances assigned to specific host servers~\cite{mackey_fly_machines_2022} that are automatically scaled up and down based on incoming events.
If the host server of the machine is resource-constrained, the machine can not be started, so that users need to over-provision machines, leading to additional fixed cost~\cite{mackey_fly_machines_2022}.
Users can manually check the status of an instance's worker host via the CLI and check the regional capacity for further instances to decide on the number of machines they should deploy.

\paragraph{Sandboxing}
Fly.io uses Firecracker~\cite{Agache_2020_Firecracker}, the same isolation mechanism as Lambda.

\paragraph{Locality}

Fly.io has multiple regions, each comprising multiple datacenters~\cite{website_flyioRegions}.
Incoming requests are forwarded to the closest machine.
Fly.io assigns every machine a unique private IPv6 address in an automatically configured Virtual Private Network~(VPN), so that it is possible to communicate from one machine to machines of the same or even other apps from the same organization\cite{ptacek_wireguard_2022}.
This is slightly different from the usual serverless programming model, where function instances are not directly addressable.

\paragraph{Configurability}

Machines can be configured to use \emph{shared} or \emph{performance} vCPUs.
A Machine configured with one shared vCPU can run \qty{5}{\milli\second} of each \qty{80}{\milli\second} period, while a performance vCPU uses \qty{100}{\percent} of a period. 
If the Machine does not exhaust this quota for some periods, the CPU time can be accumulated and used at once in a burst \cite{website_flyioCPU}.
Machines can be configured with up to 8 shared vCPUs or up to 16 performance vCPUs. 
RAM can be configured between \qty{256}{\mega\byte} and \qty{128}{\giga\byte}, depending on CPU setting \cite{website_flyioPricing}.
Machines are configured with a hard limit and soft limit for instance concurrency.
Requests will preferably route to instances that are below their soft limit, but can be routed up to the hard limit \cite{website_flyioLoadBalancing}.

\paragraph{Pricing}

Users pay for the execution duration depending on the configured resources.
Additionally, every Machine always incurs storage cost even while turned off~\cite{mackey_fly_machines_2022}.
Fly.io is the only platform where users \emph{need} to reserve enough resources to handle all load spikes and pay for these resources even if they are unused.

\subsubsection{Function Instances}

Unlike other platforms, instances need to be created ahead of time.
The worker fleet used by Fly.io all run a daemon called \emph{Flyd} to manage their locally running machines.
Flyd itself is an append-only log-based key-value store written in Go~\cite{website_flyioScheduler}.
In every region, a server called \emph{Flaps} manages all local machines by communicating with their Flyd \cite{website_flyioScheduler}.
To discover worker hosts with free capacity, \emph{Flaps} uses \emph{Corrosion}~\cite{website_githubCorrosion}, a gossip-based service discovery system.
Whenever a machine is created, Flaps queries all machines in its target region and places it based on capacity. 
When creating a machine, the worker needs to pull the image from the registry located where the function was initially created.
This leads to high variability in instance creation times, but speeds up cold starts~\cite{website_flyioRegistry}.
Before, instances were scheduled via HashiCorp's Nomad and registered in a strongly consistent centralized database in Virginia, USA, managing all platform state centrally with HashiCorp's Consul~\cite{website_flyioMachinesBlog, website_flyioConsul}.
This led to multiple issues with reliability and availability~\cite{website_flyioBadReliability, website_flyioConsul, website_flyioScheduler}.
With Flaps, instance creation is managed in a decentralized way and state is not strongly consistent due to the gossip-based state propagation approach \cite{website_flyioScheduler, website_flyioConsul}.

\subsubsection{Request Workflow}

Functions are assigned a custom subdomain that resolves to an Anycast address that is announced by all edge nodes.
This way, requests are routed to the closest available edge server, from which they are load balanced between running instances based on distance and load~\cite{website_flyioLoadBalancing}.
Based on the soft and hard limits on concurrency, the request will be forwarded to an instance in the region with capacity.
If there are no instances below their soft limit, a new instance will be started in the background.
In case there is no instance in the origin region, the request will be forwarded to another region with availability.
If all instances globally are above their hard limit, the request will be held until an instance is available, or error out after an undocumented timeout.
Unlike other platforms, Fly.io allows for instances to be addressed directly by setting the instance ID as HTTP header. 
When two instances within a region are below their soft limit, one is stopped, which is done in the order of minutes.

\subsubsection{Cold Starts}
The image is loaded to a host machine when the function instance is created, so that starting an instance comprises booting a Firecracker microVM and waiting for it to initialize~\cite{website_flyioDockerWithoutDocker}.
To reduce cold start latency, images that are scaled down can optionally be \emph{suspended} instead of stopped completely.
In this case, a snapshot of the memory is stored on the host server which execution can be resumed from during a scale up.
The platform does not guarantee that a snapshot will be used, as instances might migrate to other hosts without migrating the snapshot, requiring a normal cold start \cite{website_flyioSnapshots}.
To prevent cold starts, it is also possible to keep a specific number of instances running in the region the function was created in.

%% file: sections/5_benchmark_arch.tex
\section{Platform Performance}
\label{sec:exp_design}

\begin{table*}
    \begin{tabularx}{\linewidth}{X l p{1.25cm} S p{1.25cm}}
        \toprule
        \bf{Platform} & \bf{Languages} & \bf{RAM} & \bf{vCPU} &\bf{Arch.} \\
        \midrule
        AWS Lambda                              & JavaScript$^\alpha$, Go & \qty{128}{MB}   & 0.08   & ARM,x86 \\
                                                &                         & \qty{512}{MB}   & 0.3    & ARM,x86 \\
                                                &                         & \qty{1024}{MB}  & 0.58   & ARM,x86 \\ \addlinespace[1pt]
        Google Cloud Run Functions (CRF)        & JavaScript$^\alpha$, Go & \qty{128}{MiB}  & 0.083  & x86 \\
                                                &                         & \qty{512}{MiB}  & 0.333  & x86 \\
                                                &                         & \qty{1024}{MiB} & 0.664  & x86 \\ \addlinespace[1pt]
         Oracle Cloud Infrastructure Functions   & JavaScript$^\alpha$, Go & \qty{128}{MB}   & 0.124   &  ARM \\
        (OCI)                                   &                         & \qty{512}{MB}   & 0.5     &  ARM \\
                                                &                         & \qty{1024}{MB}  & 1       &  ARM \\
                                                & JavaScript$^\alpha$, Go & \qty{128}{MB}   & 0.249   &  x86 \\
                                                &                         & \qty{512}{MB}   & 1       &  x86 \\
                                                &                         & \qty{1024}{MB}  & 1.5     &  x86 \\
        \midrule
        Cloudflare Workers (CW)                 & JavaScript$^\beta$      & \multicolumn{3}{l}{not configurable} \\ \addlinespace[1pt]
        Deno Deploy (DD)                        & JavaScript$^\beta$      & \multicolumn{3}{l}{not configurable} \\ \addlinespace[1pt]
        Fastly Edge Compute (FEC)               & JavaScript$^\beta$      & \multicolumn{3}{l}{not configurable} \\ \addlinespace[1pt]
        Fly.io                                  & JavaScript$^\alpha$, Go & \qty{256}{MiB}  & 0.0625 & x86 \\
                                                &                         & \qty{512}{MiB}  & 0.125  & x86 \\
                                                &                         & \qty{1024}{MiB} & 0.25   & x86 \\ \addlinespace[1pt]
        \bottomrule
    \end{tabularx}
    \caption{
        Platform configuration for the experiments.
        If possible, we deployed different resource configurations to measure their impact on qualities.
        We use JavaScript as common runtime between all platforms and Go if the platform supports it.
        \footnotesize{
            $^\alpha$via a full JavaScript runtime.
            $^\beta$via V8 isolates.
        }
    }
    \label{tab:experimentOverview}
\end{table*}

To better understand the real-world implications of the different platform architectures and verify the expected behavior based on their architecture, we benchmark their performance across different dimensions.
We run several benchmarks on all platforms, focusing on microbenchmarks that enable a fair comparison between platform providers.
An overview of the function configuration for all experiments is shown in \cref{tab:experimentOverview}.
We focus on experiments measuring the following qualities of serverless platforms, that have been identified as critical factors in previous research~\cite{Eismann_2021_State} and are closely related to the overall architecture of the platform.

\paragraph*{Warm Latency}

To measure the response time of warm functions, we deploy a function that immediately returns an empty result.
We call the functions with an Inter-Arrival Time~(IAT) of three seconds for one minute, and then wait five minutes before repeating.
This experiment measures client-side latency, the time inside the function when it was started, and whether it is a cold start.
Due to possible clock drift between the client and platform, we focus on the client-side latency.
If the platform is regional, we configure it to run inside the cloud platform's region closest to Frankfurt, Germany, where the benchmarking client runs, to minimize network latency.
This does however mean that small latency differences between platforms could be caused by the network in-between the benchmarking client and the cloud platform.

\paragraph*{Cold Start Latency}

The goal of this experiment is to measure how long cold starts take.
While the warm latency experiment also has cold starts, in this experiment, the functions are called once every 15 minutes to ensure higher frequency of cold starts.
Otherwise, the configuration is the same.

\paragraph*{Geo-Distributed Workloads}
Newer platforms focus on simplifying geo-distributed serverless deployments, which we benchmark by calling the functions from clients distributed all over the world and measuring the request response latency.
We use the infrastructure provided by Grafana Cloud which is build on top of AWS to send these requests from 20 different regions~\cite{website_grafanaCloudRegions}.
CW, Fly.io, FEC, and DD support deploying the same function to multiple regions and route requests to the closest available region.
We configured them to use all possible regions, and left the other platforms in the same regions as before.
The goal of this experiment is directly comparing different platform's performance, so instead of using all resource configurations, we only use the smallest available JavaScript function.
We call the functions every three days at 12:00 CET with an IAT of 1 second for 4 minutes.

\paragraph*{Performance Consistency}
Previous research has shown that serverless platforms do not offer consistent performance inside function instances with varying daily and weekly performance~\cite{schirmer2023nightshift}.
Due to the pay-per-use model of serverless, this means that when platforms perform slower, they also get to charge users more.
To measure the variability, we deploy one function that runs a CPU heavy task, and another function that writes and reads a file from disk.
We subject all platforms to the same CPU and I/O microbenchmarks and measure the time it takes for them to complete the task.
In particular, for CPU performance we calculate the sum of the products of the square root and the sinus value of a running index \emph{i} in 100 iterations.
To measure I/O performance, we create a string of \SI{10}{KB} and write and then read it from the file system.
We then call them twice every 15 minutes for, on average, one cold and one warm start over multiple weeks.

\paragraph*{Elasticity}
Elasticity measures how fast the platform can react to load changes from specific functions~\cite{book_bermbach2017_cloud_service_benchmarking}.
From a client perspective, the serverless platform changes observably with latency spikes during load bursts.
To evaluate how the platform responds to rapidly changing load patterns, we linearly increase the number of requests per second (RPS) from one to 20 in nine seconds, then stay at 20 RPS for one second, and then ramp down to zero in one second.
This load pattern is unlikely to impact \emph{global} platform load, as these platforms serve millions of requests per second~\cite{Brooker_2023_OnDemand,sahraei2023XFaas}.
We opted for this pattern to keep cost at acceptable levels~\cite{book_bermbach2017_cloud_service_benchmarking} and investigate whether rapid per-user changes can influence performance for this user.
We measure the request response latency for all requests and repeat this experiment every 15 minutes.

\paragraph*{Data Transfer}
Serverless workflows frequently transfer large amounts of data between functions.
This can be solved by integrating with cloud storage services such as S3 or by directly transferring data between functions, which both are a significant source of latency in serverless applications~\cite{Copik_2021_Sebs,Ustiugov_2021_Stellar}.
We run experiments to measure how long it takes to directly transfer data between two functions (producer and consumer) in the same region.
We measure the duration from the invocation of the consumer by the producer until the entire message is parsed.
All platforms are configured with the smallest resource configuration and 1 MB random text is used as payload.
We repeat this once every 15 minutes.

\subsection{Platform Configuration}

Some platforms enable users to choose the amount of resources that function instances have access to.
In that case, we deploy functions with different configurations with the goal of detecting \emph{whether} resources impact the qualities.
An overview of the configuration per platform is shown in~\cref{tab:experimentOverview}.
We configure Lambda to use 128, 512, and 1024 MB RAM, with the platform assigning 0.08, 0.3, and 0.58 vCPUs.
For CRF, we configure the same amount of RAM and use the minimum possible number of vCPUs.
For OCF, we reuse the same RAM configurations with both ARM and x86 instances.
For Fly, we configure machines with \qtylist[list-units=single]{256;512;1024}{MiB} RAM.
We configure the smallest instance with the minimal possible vCPU and double the vCPU configuration for the other configurations.
Note that the way JavaScript is supported changes between platforms --- some use higher-level isolation mechanisms such as V8 isolates, others use lower level mechanisms such as microVMs.

%% file: sections/6_benchmark_results.tex
\section{Experiment Results}
\label{sec:results}

For warm starts, the consistency of function performance and the frequency of cold-starts are important, as they have a major impact on total latency and cost.
Across all experiments, we observe a clear split between edge--oriented (CW, DD, FEC, and Fly.io) and regional platforms (Lambda, CRF, OCF): Edge platforms typically deliver substantially lower median latencies and tighter latency distributions, while regional platforms tend to be slower with more variance.
This clustering is strongest for warm-start and data-transfer performance, whereas cold-start behavior is both more platform-specific and markedly less stable overall.
Cold starts introduce large variance and dominate tail latencies; the difference between cold and warm starts differs widely by platform (\cref{fig:res:cold-performance}).
Implementation and configuration choices also matter: Compiled languages (Go) and larger resource configurations consistently improve latency, while the interpreted JavaScript slows down warm as well as cold starts.
For geo-distributed workloads, regional platforms have increased latency with increased distance to the benchmarking client, whereas edge platforms remain low-latency and comparatively uniform across locations.
Elasticity dynamics largely reflect the combination of warm-start baselines and cold-start delays we measured in their respective experiments.
Finally, we observe occasional time-bounded performance degradations on individual platforms.
The remainder of this section evaluates each experiment in turn and highlights the platform- and setup-specific effects that explain these aggregate patterns.
We focus on two main descriptive statistics: The median ($\mu$) latency to measure the central tendency, and Pearson’s~r as a measure of linear correlation.
Pearson’s~r is a normalized covariance with values between -1 and 1, with values closer to the extreme indicating a stronger linear relation.
All experiments ran from 2025-11-13 to 2026-01-05, resulting in \SI{38165688}{} recorded function calls.

\begin{figure}
    \centering
    \includegraphics[width=\linewidth]{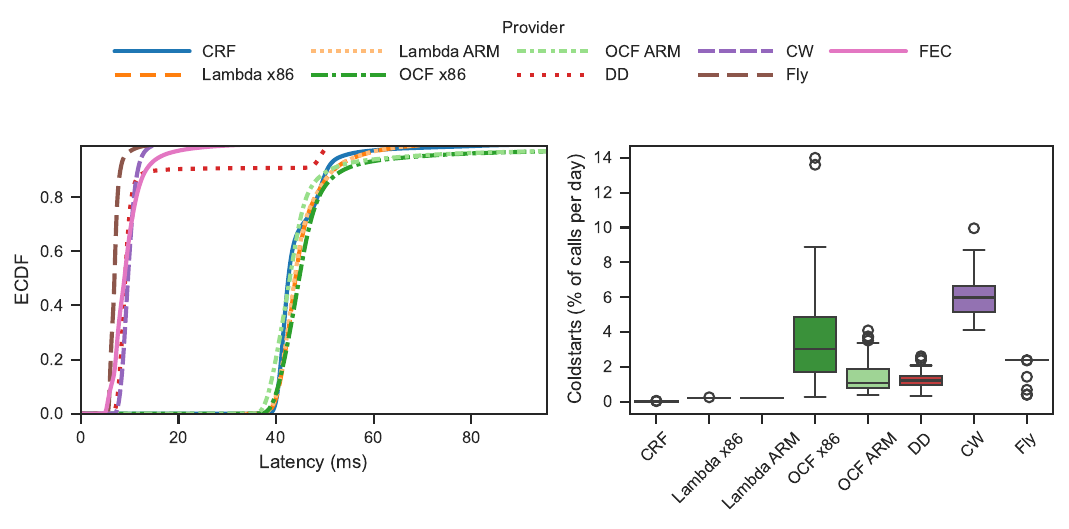}
    \caption{\emph{Warm start performance.} The left ECDF shows two distinct clusters that both have similar latency distributions: edge-oriented (DD, CW) with ${\mu}{\approx}$\SI{10}{ms}, and regional platforms (OCF, CRF, Lambda) with ${\mu}{\approx}$\SI{40}{ms}.
    In the ECDF, the slowest \qty{5}{\percent} are not shown.
    The bar chart on the right visualizes the relative frequency of cold starts per day in the warm start experiment.
    While most platforms have a low unexpected cold start frequency of ${\leq}$\qty{2}{\percent}, OCF, DD, and CW are outliers.
    }
    \Description{On the left, an ECDF shows that DD, CW, and Fly.io follow a steep slope at less than \SI{10}{ms} until the 80th percentile. Afterwards, DD has a plateau and only continues to climb after \SI{40}{ms}, while the others continue rising.
    The regional providers follow a normal distribution centered around \qty{40}{ms}, with a long plateau after the 90th percentile.
    The right Bar plot shows that all platforms except for Lambda ARM and CW have a cold start percentage under \qty{2}{\percent}, while Lambda is at \SI{3}{ms} median with a 25th percentile around \SI{2}{ms} and its 75th percentile at \SI{5}{ms}.
    CW has its bar between \qtyrange[range-units=single]{5}{8}{\percent}.
    }
    \label{fig:res:warm-performance}
\end{figure}

\subsection{Warm Latency}
\label{subsec:warm-latency}

For the warm latency experiment, the consistency of function performance and the cold start incidence are important, as these factors influence user experience and cost.
Our results largely show expected behavior with function performance and coldstart incidence being consistent within each platform.
However, there is a noticeable difference between the regional and edge-oriented platforms.
Edge platforms show warm median latencies around \SI{8}{ms}, while regional platforms have median latencies around \SI{40}{ms}.
While this is partially caused by network latency between our client and the respective platform, it shows that edge-oriented platforms do not require optimizing client placement to achieve low latencies.
Notably, the relative frequency of cold starts does not follow the same two clusters.
\Cref{fig:res:warm-performance} shows that Lambda and CRF exhibit the lowest median coldstart percentages per day, while the two highest cold start frequencies are from OCF and CW.
We also observe that changes in chip architecture do not lead to performance differences of the same magnitude across platforms.
While Lambda shows only \SI{0.1}{ms} differences between ARM and x86 median latencies, OCI exhibits a noticeably larger gap: median latency differs by \SI{2}{ms} between architectures, and the variability changes between chip architecture, with aggregated standard deviations of \SI{287}{ms} on x86 (\qty[print-implicit-plus]{315}{\percent}) and \SI{402}{ms} on ARM (\qty[print-implicit-plus]{360}{\percent}).
OCI also shows significantly higher tail latencies than all other platforms, with its 99th percentile reaching up to \SI{811}{ms}, while maintaining an interquartile range similar to the other platforms.
Since these tail latency events are likely caused by resource exhaustion in the workers, this indicates to us that the ARM-side of the OCI operates under higher resource utilization.
If these effects had been caused by the underlying platform \textit{architecture}, they would be the same for x86 as well.
None of the platforms showed notable differences between resource configurations.

\subsubsection{CRF}
\label{subsec:results:warm:crf}
For CRF, we observe two significant intermittent performance degredations during the timeframe of the experiment.
Between the 24th of November and the 5th of December, median latency increased by approximately \qty{20}{\percent}.
Additionally, between the 21st and 24th of December, median latency increased by approximately \qty{10}{\percent}.
During the first performance degradation, the variance of execution durations stayed low, and platform performance returned to normal afterwards.
The second degradation, which happened during the Christmas break, coincided with a higher variance of execution durations and was reverted quicker.
Unexpected events like these make application performance less predictable, introducing additional complexity for users.

\subsubsection{DD}

The median latency is comparable to the other edge platforms, however its consistency is worse. DD’s 95th-percentile latency reaches up to \SI{49}{ms}, a \qty{444}{\percent} increase over its median, whereas Fastly Edge Compute and Cloudflare Workers show only an \tRange{80}{90}{\percent} increase from the median to the 95th percentile.

\subsection{Cold Latency}
\label{subsec:cold-latency}

\begin{figure}
    \centering
    \includegraphics[width=\linewidth]{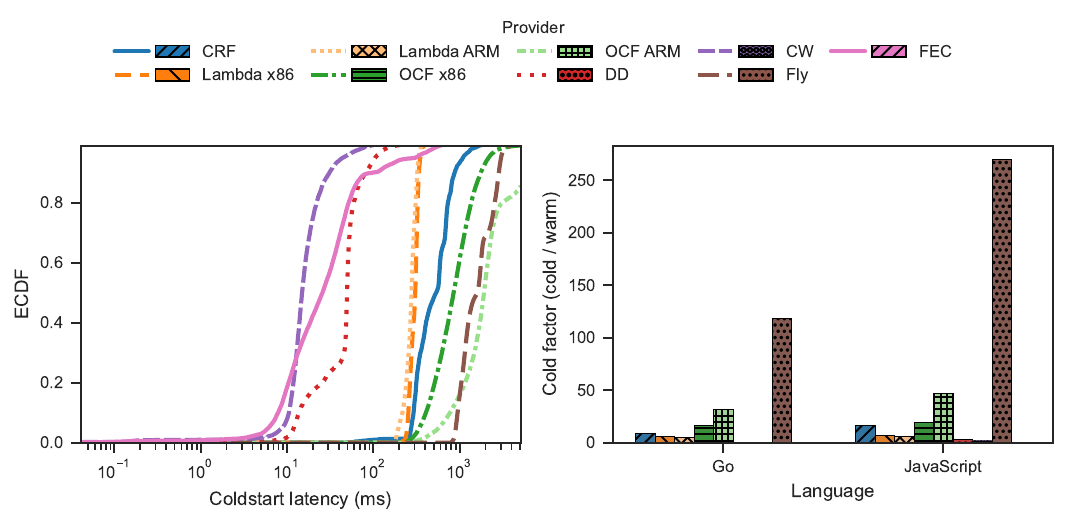}
    \caption{\emph{Cold start performance.}
    The ECDF on the left shows that the two latency clusters observed during warm start experiment are less pronounced for cold starts, with higher spread across providers.
    The box plot on the right displays the cold factor $\frac{\mu_{\text{cold}}}{\mu_{\text{warm}}}$:
    Fly consistently shows the highest penalty for cold starts, whereas Deno Deploy and Cloudflare exhibit near-zero cold factors, indicating minimal overhead compared to their warm start performance.
    }
    \label{fig:res:cold-performance}
\end{figure}

In contrast to warm starts, cold start latencies are less stable and exhibit higher variance.
As a result, median cold start latencies --- showing larger discrepancies across platforms --- become the primary metric of interest.
To compare overall platform performance across both operational modes, we use the \textit{cold factor} for which we divide median coldstart latency by median warmstart latency for each provider.
This provides an intuition for the difference between the behavior modes to evaluate overall platform performance:
$CF = \frac{\mu_{\text{cold}}}{\mu_{\text{warm}}}$.
Compared to the clustering from the warm start experiment, cold latencies show a more dispersed distribution for each platform class, as visible in~\cref{fig:res:cold-performance}.
Additionally, there are two language effects.
First, JavaScript cold starts are at least \qty{18}{\percent} slower than Go across all platforms.
Second, the gap between cold-start and warm-start latencies is larger for JavaScript than for Go, which is reflected in the \textit{cold factor} in~\cref{fig:res:cold-performance}.
The resource configuration has no effect on most platforms except Fly.io and CRF.

\subsubsection{CRF}

Although CRF shows consistent performance overall, there are minor differences between resource configurations:
increasing resources also improves latency.

\subsubsection{Lambda}

ARM-based functions display marginally better median latencies than x86-based functions (\SI{270}{ms} and \SI{295}{ms}) but in exchange show decreased consistency as exhibited in an increased Interquartile Range (IQR) (\SI{24}{ms} and \SI{41}{ms}) and standard-deviation (\SI{30}{ms} and \SI{45}{ms}).

\subsubsection{OCF}

For Oracle's platform, this effect is substantially stronger.
On ARM, the median latency is \SI{1917}{ms}, compared to \SI{839}{ms} on x86 (\qty[print-implicit-plus]{129}{\percent}).
Dispersion measures show the same trend:
the IQR is \SI{2121}{ms} on ARM versus \SI{657}{ms} on x86 (\qty[print-implicit-plus]{223}{\percent}), and the standard deviation is \SI{3557}{ms} on ARM compared to \SI{2366}{ms} on x86 (\qty[print-implicit-plus]{50}{\percent}).
Moreover, Oracle exhibits the highest 99th-percentile latencies of all platforms on ARM (\SI{9740}{ms}).

\subsubsection{Fly}

Fly exhibits the widest discrepancy between median cold- and warm start latency.
Although, the platforms warm starts took a median of only \SI{7}{ms}, its coldstarts show a median latency of \SI{1523}{ms}, an increase of \qty{21657}{\percent}.
Fly.io is the only platform that shows \emph{worse} performance with higher resource allocations.

\begin{figure}
    \centering
    \includegraphics[width=\linewidth]{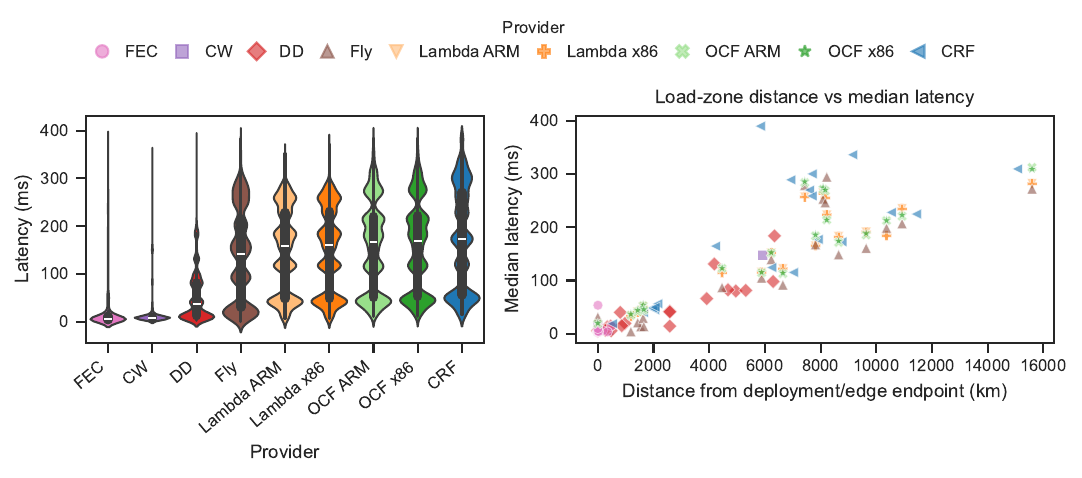}
    \caption{\emph{Global latency performance across providers.} The violin plot on the left show that regional providers exhibit wider, multimodal latency distributions reflecting the varying geographical distances between clients and deployed functions, while edge-oriented providers remain tightly concentrated near zero.
    The scatter plot on the right confirms a strong distance--latency correlation for regional providers, with edge-oriented providers clustering near the origin regardless of client distance.
    }
    \Description{
    The left violin plot has latency on the y axis and the x axis is the providers.
    The regional providers all have multiple bumps between 0 and 400ms, while FEC, CW, and DD are all around 10ms.
    The right plot has the distance on the x axis and the median latency on the y axis.
    It shows a clear straight line on which most providers lie, but also some outliers towards the top, mostly with CRF and OCF.
    }
    \label{fig:res:global-performance}
\end{figure}

\subsection{Geo-Distributed Workloads}

Unsurprisingly, edge platforms outperform regional platforms for geo-distributed workloads.
While the latter shows multimodal latencies when analyzing the data at a global scale, edge-platforms show a uniform distribution with equal latencies across all tested locations (\cref{fig:res:global-performance}).

CRF, Lambda, Fly and OCF show median global latencies between \SI{142}{ms} and \SI{173}{ms} as well as interquartile ranges of similar widths between \SI{169}{ms} and \SI{215}{ms}.
The multimodal latency distributions shown in~\cref{fig:res:global-performance} align with the geographic separation between the load zones and the region where the function is deployed.
In~\cref{fig:res:global-performance}, each dot or cross represents the median latency observed for a single load zone, revealing a strong distance–latency relationship (Pearson’s~r~=~0.845) for regional platforms.

FEC and CW exhibit consistently low latencies with unimodal distributions, achieving median latencies of \SI{4}{ms} and \SI{7}{ms}, respectively.
Their interquartile ranges are narrow (\SI{8}{ms} for Fastly and \SI{3}{ms} for Cloudflare), indicating little variability across load zones.
This unimodality is consistent with the platforms’ dense edge footprints and the proximity between load-generating functions and nearby points of presence, which minimizes client–server distance.
In the right plot of~\cref{fig:res:global-performance}, this is reflected by the tight clustering of CW and FEC data points near the origin, alongside a strong distance–latency correlation (Pearson’s~r~=~\SI{0.934}{}).
Interestingly, while DD is also classified as an edge-oriented platform, its latency distribution remains multimodal (albeit less pronounced than for regional platforms).
This observation is in line with the corresponding distance–latency relationship, with distances extending to roughly \SI{7000}{km} and a correlation of r = \SI{0.887}.
We attribute this behavior to DD’s comparatively smaller Point of Presence (PoP) footprint, which leaves some load zones without a closely matching deployment region.

\subsection{Performance Consistency}

\begin{figure}
    \centering
    \includegraphics[width=\linewidth]{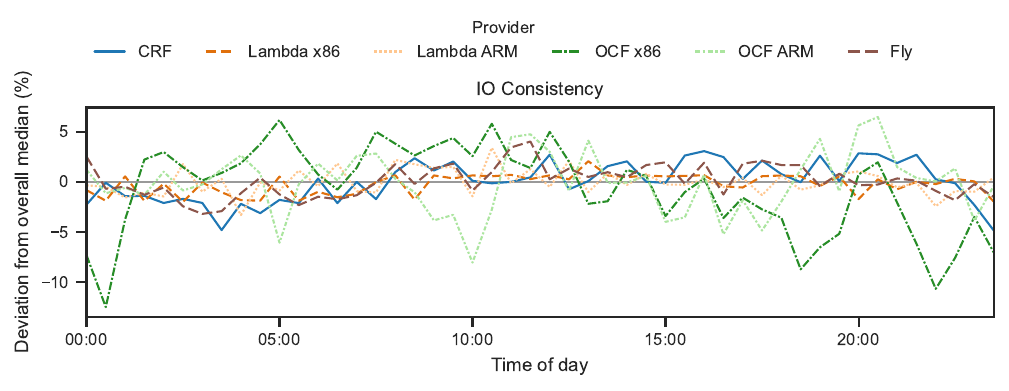}
    \caption{\emph{I/O consistency.}
    The plot shows the deviation from the overall median I/O consistency for providers supporting file system interactions, depending on time of day.
    OCF and Fly, both using OCI images, consistently achieve the lowest absolute I/O latencies, but OCF has the highest variance.
    Lambda has average absolute values with very low variance, and CRF exhibits the highest I/O latencies with medium variance.
    }
    \Description{Line plot showing the relative performance deviation from the median during the 24 hours of a day.
    There is no real trend, but most providers are slower during the day.}
    \label{fig:res:cpu-io-consistency}
\end{figure}

We investigate two classes of performance consistency:
(i) how consistent the CPU is and (ii) how I/O performance behaves (\cref{fig:res:cpu-io-consistency}).
Although we observe that calculation times --- which we use to measure CPU consistency --- are stable across platforms when comparing medians, they are not stable internally for most platforms.
That is, while calculations take a median of \SI{20}{ms}, the aggregated IQR over all platforms is \SI{80}{ms} wide.
Contrarily, I/O operations take more time in the median at \SI{490}{ms} but show comparatively more stable performance with an IQR of \SI{103}{ms}.
We noticed no patterns in cross-platform CPU performance.

\subsubsection{CRF}

CRF exhibits comparatively high consistency and low calculation times for the CPU evaluation with a median of \SI{20}{ms} and an IQR of \SI{30}{ms}.
Contrarily, it displays the slowest I/O operation times across all platforms at a median of \SI{2002}{ms} and a high variation with an IQR of \SI{401}{ms}.

\subsubsection{Lambda}

Lambda has the opposite behaviour:
it has slow performance for CPU tasks on x86 (µ=\SI{150}{ms}; IQR=\SI{150}{ms}) and ARM-architecture (µ=\SI{220}{ms}; IQR=\SI{220}{ms}), but shows consistent I/O behavior across architectures with comparatively lower median operation times of approximately \SI{900}{ms}.

\subsubsection{OCF}

Median computation times are comparable across architectures for OCF ($\mu_{\mathrm{ARM}} = 10\,ms$; $\mu_{\mathrm{x86}} = 20\,ms$), but variability differs markedly.
On ARM, the IQR is \SI{10}{ms}, whereas on x86 it widens to \SI{120}{ms}.
For I/O, we observe the opposite pattern:
while the IQRs are similar across architectures, the median I/O time diverges substantially relative to the corresponding median computation times.

\subsection{Elasticity}

\begin{figure}
    \centering
    \includegraphics[width=\linewidth]{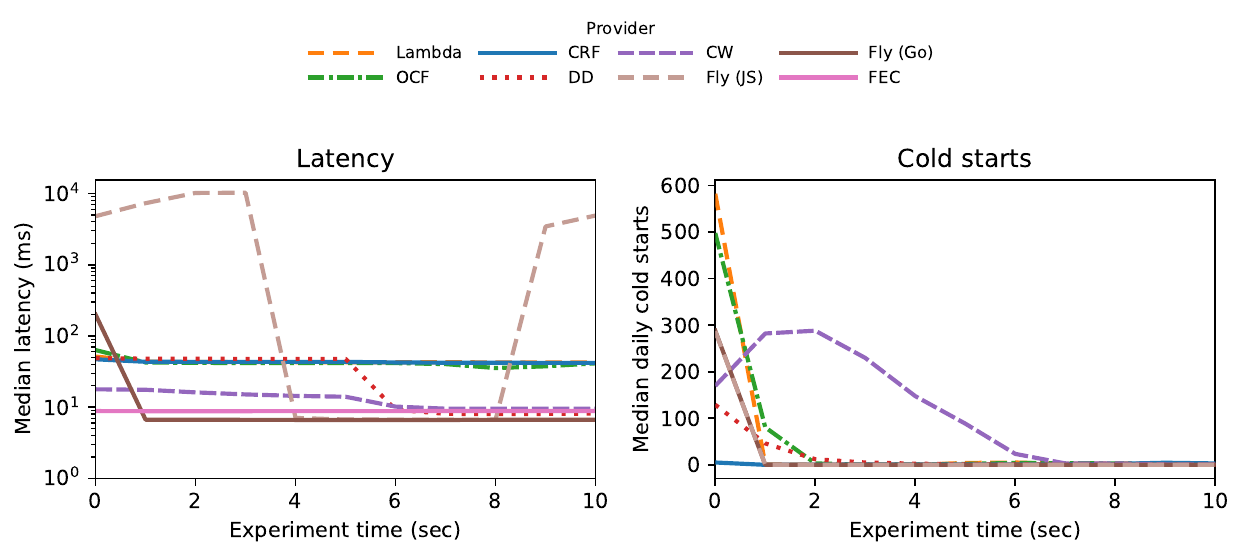}
  \caption{\emph{Median latency and cold start occurence in the elasticity experiment.}
  The median per-second latency over the course of the experiment (left plot) uses a logarithmic scale as providers have vastly different performance.
  Latency peaks initially due to cold-start execution and decreases as platforms transition to mostly serving requests from warm instances.
  The right plot shows the median number of cold starts experienced during a day the experiment ran.
  While most platforms have all cold starts in the beginning, CW has its peak after two seconds.
  The cold start trajectories for Fly.io are identical using JavaScript and Go.
  FEC is excluded from the right figure, as does not provide a way of detecting if an instance is warm or cold.
  }
  \Description{The left plot shows a mostly flat latency at 40 ms for CRF and OCF and Lambda. FEC and CW are consistently below 20ms. DD is at 50ms for the first 6 seconds and then drops to 10ms.
  }
    \label{fig:res:elas-plot}
\end{figure}

The results of this experiment reflect a combination of the effects seen in the warm-start and cold-start experiments (see \cref{subsec:warm-latency,subsec:cold-latency}).
For all platforms except Fly.io and CRF, latency traces showed no difference between different resource configurations.
As in the warm-start setting, platforms cluster into regional and edge-oriented platforms.
CRF, OCF, and Lambda show predictable elasticity dynamics:
Latency values are higher in the first seconds due to cold starts, followed by stable, lower latencies.
Although steady-state performance is stable, longer cold-start delays reduce elasticity by slowing how quickly the platform can absorb load.
FEC and CW are also broadly stable, but their traces deviate more from their median baselines and are less consistent overall;
however, shorter overall latencies, and especially shorter cold-start times, improve elasticity (see~\cref{fig:res:elas-plot}).

As the per-second cold-start rate is strongly correlated with the per-second median latency, we argue that the additional load this experiment puts the platform under does not influence its overall behavior noticeably.

\subsubsection{CRF}

In line with results from the warmstart evaluation, CRF shows few coldstarts, mostly within the first second.
Even though function configurations are largely consistent for the platform, the \SI{128}{MiB} JavaScript function displays slightly higher latencies and more dispersed coldstart incidence.
Although overall latencies behave similarly, the cold starts for the smallest configuration occur later in the experiment, indicating longer cold start times for functions with limited memory.
This is consistent with the performance we see in our cold start experiments, where latencies are higher for the lower resource configurations.

\subsubsection{Lambda}

On Lambda, performance is not only stable across all configurations but also for both CPU-architectures.
All configurations exhibit the same latency and coldstart patterns, where maxima for both metrics occur in the initial second, and stabilize afterwards.
The congruity shown in the data of the elasticity experiment corresponds to the consistency in Lambdas warm- and coldstart performance.

\subsubsection{OCF}

On x86, the \SI{128}{MB} configuration shows the highest latencies for both languages --- approximately \qty{175}{\percent} higher than the other configurations -- limiting the configuration's ability to scale efficiently under changing load.
In addition, cold-start behavior is less predictable than for the previously discussed platforms, as indicated by a markedly wider IQR within the first second.

On ARM, the \SI{128}{MB} and \SI{512}{MB} JavaScript functions show a high median latency of \SI{1200}{ms} in the initial second --- an increase of approximately \SI{380}{\percent} over the other configurations.
Otherwise, OCF's elasticity reflects the behavior observed in the cold- and warm start experiments.

\subsubsection{Fly}

Fly shows the largest elasticity gap between Go and JavaScript among all platforms.
Because JavaScript coldstarts take around \SI{3}{s}, latencies continue to rise until additional instances become available, which visibly degrades elasticity and causes peak median latencies, that exceed Go’s peak medians by \qty{2757}{\percent}.
Once all instances are started, the improvement in latency is better for smaller resource configurations, correlating with our findings from the cold start experiment.
Colds starts in Go complete in roughly one second, causing most cold-start activity to be concentrated in the first second.

\subsection{Data Transfer}

\begin{figure*}[t]
  \centering
  \begin{subfigure}[t]{0.49\textwidth}
    \includegraphics[width=\linewidth]{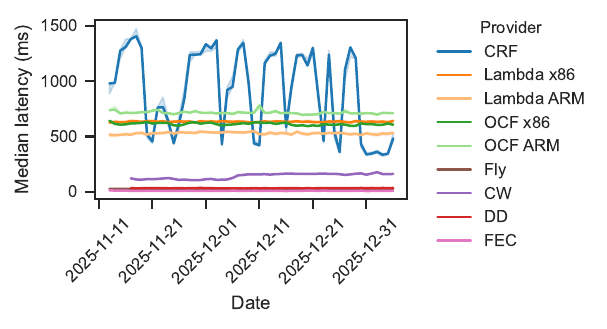}
    \caption{\emph{Daily median latencies.}}
    \Description{Transfer latencies across platforms}
    \label{fig::res:transfer-latency}
  \end{subfigure}\hfill
  \begin{subfigure}[t]{0.49\textwidth}
    \includegraphics[width=\linewidth]{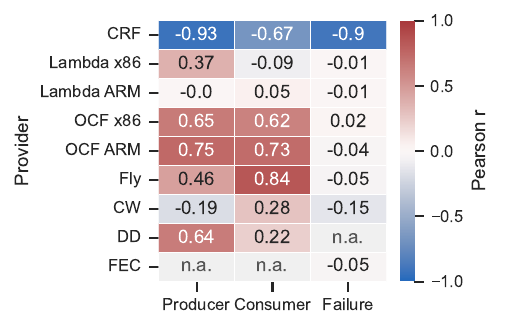}
    \caption{\emph{Correlation between transfer latency and cold start / failure rates.}
    }
    \Description{CPU and I/O-performance consistency}
    \label{fig::res:corr-latency-coldstart-failures}
  \end{subfigure}

  \caption{\emph{Transfer latency performance.} Again, a clear clustering between regional platforms and edge-oriented platforms emerges.
  The daily median latency plot on the left shows that the performance is relatively consistent for the duration of the experiment, with the regional platforms hovering around \SI{600}{ms} and the edge-oriented platforms around \SI{30}{ms}.
  CRF is the only highly variable platform.
  The correlation plot on the right shows that CRF has strong negative correlations across all three dimensions, while OCF and Fly exhibit moderate-to-strong positive correlations between latency and cold start rates on both producer and consumer sides.
  Lambda, CW, and FEC show correlations close to zero, suggesting latency is largely independent of cold start and failure behavior for those providers.
  FEC consistently crashed when using \SI{1}{MB} payload.
  To still measure platform transfer performance, we had to reduce its file size to \SI{30}{KB}.
  }
  \Description{Performance for the Transfer latency experiments}
  \label{fig:res:performance-transfer}
\end{figure*}

As shown in \cref{fig::res:transfer-latency}, transfer latencies show a clustering similar to the warm-start results:
regional and edge-focused platforms form two distinct groups.
The regional platforms achieve median transfer latencies of roughly \SI{600}{ms}, whereas edge platforms cluster around \SI{30}{ms}.
Overall, transfer performance is largely stable within each platform, with a few exceptions discussed below.
On FEC, functions always crashed when transferring \SI{1}{MB} of data, so that we reduced the data transfer to \SI{30}{KB} for FEC.
Since cold starts are typically slower but less frequent than warm starts, and because cold-start incidence correlates with transfer latency for most regional platforms (\cref{fig::res:corr-latency-coldstart-failures}), their 95th and 99th percentiles show pronounced increases in transfer latency.

\subsubsection{CRF}

\cref{fig::res:transfer-latency} shows CRF's pronounced day-to-day variability:
it has the widest IQR of all platforms (\SI{837}{ms}) and comparatively slow transfers, with a median latency of \SI{1060}{ms}.
Since cold-start incidence for both producers and consumers is positively correlated with the failure rate, and the failure rate is strongly negatively correlated with median latency, the apparent latency reduction on days with high cold-start incidence is likely a selection effect:
long-running cold starts are underrepresented in the latency distribution on those days because they fail rather than contribute to the aggregated latencies.
As the timeout we configured for functions is nine minutes, this seems to be another internal timeout that is not user-controllable.

\subsubsection{Lambda}

Transfer latencies are more stable and there are no correlations between coldstarts incidence and latency changes.
However, standard deviation ($\sigma_{x86}=372~ms$; $\sigma_{ARM}=412~ms$) and IQR ($IQR_{x86}=210~ms$; $IQR_{ARM}=249~ms$) still hint at occasional inconsistencies.
Functions on ARM-architecture show a decreased median latency by approximately \SI{100}{ms} when compared to x86-based functions.

\subsubsection{OCF}

OCF exhibits a mirrored cross-architecture pattern:
x86-based functions achieve a roughly \SI{100}{ms} lower median latency, while the standard deviation and interquartile ranges are substantially smaller than those observed for AWS Lambda.
Notably, there are strong positive correlations between producer and consumer cold-start rates and the median transfer latency for both architectures.
This is caused by Oracle’s comparatively long cold-start times (\cref{subsec:cold-latency}).

\subsubsection{Fly}

Transfer latencies on Fly are largely stable, with a median of \SI{27}{ms} and an IQR of \SI{8}{ms}.
However, on December 8 and 9, we observe a pronounced increase in variability.
This change is driven by a surge in the tail:
the 99th percentile rises from approximately \SI{200}{ms} to approximately \SI{2700}{ms}.
In contrast, the 95th percentile does not exhibit a single sustained shift;
instead, it becomes bimodal after these dates, alternating between roughly \SI{200}{ms} and \SI{2200}{ms}.

\subsubsection{CW}

CW exhibits mostly consistent transfer latencies, with a standard deviation of \SI{42}{ms}.
Nevertheless, the daily median increases on December 6 and 7, shifting from approximately \SI{120}{ms} to \SI{160}{ms}.
Moreover, it shows substantially higher transfer latencies than the other edge-oriented platforms (FEC, DD), despite matching or outperforming them in our other experiments.

%% file: sections/7_discussion.tex
\section{Discussion and Future Work}

Deployment models are arguably the clearest factor differentiating first- and second-generation FaaS platforms.
The earlier serverless platforms up to about 2019 are inherently regional platforms.
They exhibit a unique architectural style: they use microVMs or containers as isolation mechanisms, and heavily focus on reliably routing requests to suitable function instances.
Lambda, CRF, and OCF are clear members of this group.
After 2019, a new paradigm has emerged, with CW, DD, and FEC as its key representatives: Rather than requiring users to restrict their functions to a single region, workloads are deployed and globally replicated by default.
These platforms achieve this high degree of replication through the use of higher-level, process-based isolation mechanisms using V8 isolates, drastically reducing the memory and storage footprint of function instances.
Fly.io sits in between those two groups.
Introduced in 2020, it follows a different resource model than all other platforms, relying on the older-style microVM isolation mechanism with high resource limits while also being edge-focused.
However, our evaluation shows that they are not able to keep up with the performance of neither the old nor the newer platforms.

\paragraph{Simplicity vs. Flexibility}

A noticeable change in prioritization between the two generations is a shift from flexibility to simplicity.
Over the years, older platforms have evolved to provide users with an increasing number of configuration options for function behavior.
In first-generation platforms, users can finely configure the resources of their instances. 
This configurability offers the flexibility needed to adapt the paradigm to a wide array of use cases such as resource-intensive machine learning workloads~\cite{yang_infless_2022,werner2024serverlessBigData}.
The second-generation platforms take a different approach: They prioritize a simple deployment model optimized for workloads comprising short-lived executions at a global scale, aiming to reduce latency as far as possible through lightweight virtualization and client proximity.
For instance, CW and DD primarily offer JavaScript and TypeScript as their language runtime, do not allow users to configure memory, and have strict limitations on execution duration.

\paragraph{Performance vs. Security}

The separation into first and second generation can also be observed when differentiating virtualization technologies and their impact on performance.
While first generations use complex virtualization at the OS level through microVMs or containers, the new generation uses process-based isolation.
The exact implementation of the isolation mechanism differs between platforms:
While OCF uses standard OCI containers without additional middleware to improve performance or security, CRF employs gVisor. 
This enhances the security and isolation guarantees of its container-based virtualization, albeit with potential performance trade-offs.
Similarly, Lambda achieves highly consistent performance in all experiments using Firecracker, indicating well-tuned background infrastructure and load balancing. 
Conversely, as seen in our experiment results, Fly.io cannot provide those same guarantees, even though they also use Firecracker.

Although process-based isolation enables a simpler architecture and improves performance by reducing memory footprints and cold-start latencies, it also makes the newer platforms more susceptible to security risks as it offers a larger attack surface for side-channel attacks~\cite{marin_serverless_2022}.
This model also restricts what can be done inside the execution environment, as system resources are heavily constrained and many kernel functionalities are inaccessible.
On the contrary, more complex and safe virtualization technologies have a higher overhead, negatively impacting performance in warm- and coldstart scenarios as well as reducing elasticity.

\paragraph{Information on Platforms}

Because the platforms we focused on are all closed-source commercial FaaS offerings, the goal of this paper was to aggregate publicly available information on them and evaluate them comparatively. 
However, as the platforms do not publish their internals to protect their competitive advantage, not all information is publicly available, and certain architectural details remain unknown.
Furthermore, it is always possible that platform providers have updated their internal system designs without publishing the corresponding documentation.
Because FaaS platforms, especially of the second generation, are still actively developed, it is likely that some information is already outdated.
Yet, we argue that the approach of collecting all available information and enhancing it with empirical evaluation data is the only viable approach to studying closed-source platforms, which are heavily used in both industry and academia.

Another limitation of this study is its longevity.
The platforms discussed by Wang et al. in 2018~\cite{Wang_2018_Peeking} were from the biggest cloud providers at the time and all still exist.
However, in a similar study conducted in 2017, Lynn et al.~\cite{Lynn_2017_ReviewAllEnterprise} published a preliminary review of seven enterprise serverless platforms; of those, only three remain operational today.
This is an indicator of the uncertainty regarding whether the platforms studied in this paper will still exist in a few years.
Notwithstanding, they serve as a valuable snapshot of the direction the overall industry is moving in, even if individual platforms ultimately fail.

\paragraph{Experiment Scope}

In our experiment design, we focused on benchmarking the same dimension across platforms.
To achieve this, we focus on microbenchmarks.
Including application benchmarks that covers a more realistic use case would introduce too many confounding factors and ultimately obfuscate the platform-specific performance characteristics we want to identify~\cite{Copik_2021_Sebs,paper_grambow2021_befaas,yu_characterizing_2020}.
For this reason, our experiments do not include a cost analysis, as our microbenchmarks do not constitute realistic load patterns, and therefore a cost analysis based on them would misrepresent real-world billing behavior.
For instance, most of our functions are empty and therefore have minimal execution durations, which does not represent a realistic FaaS workload, especially since platforms often bill execution time at a very fine granularity.
As a rough estimate, the total monthly cost for all experiments was \qty{\sim 30}{USD} on Fly.io, \qty{\sim 0.5}{USD} on CRF, and \qty{\sim 15}{USD} on Oracle.
All other platforms incurred no cost, as our usage fell into the free tier.
Building on this work, future research could focus on identifying use cases for the different generations of platforms, building representative applications, and measuring them.

\paragraph{Platform Extensions}

Another direction for future work is the investigating how platform extensions can be integrated into the different platform architectures.
For example, stateful FaaS workloads—such as AI inference pipelines, streaming, and batch processing—are increasingly prevalent~\cite{yang_infless_2022,paper_pfandzelter2022_streamingfunctions,muller_lambada_2020}, yet it remains unclear how well the architectural trade-offs examined in this paper translate to such scenarios. 
In particular, investigating whether specific virtualization, deployment, or control plane designs are inherently better suited for new use cases is a promising direction, as it may extend the requirements for a good FaaS platform architecture.

\paragraph{Double Billing}

The double billing problem~\cite{Baldini_2017_Trilemma}, a core issue of serverless research in the last years, has essentially been solved by the new platforms.
Since these platforms measure active CPU cycles rather than wall-clock execution time, functions awaiting I/O or other function responses incur no additional compute costs~\cite{cf2020workers}. 
Serverless workflows have historically struggled with cost-efficiency due to per-request billing~\cite{schmid2025sebsFlow}, a barrier CW has solved by simply not charging for internal requests executed as part of a workflow.
As theoretical pricing models are often based on the traditional first-generation architecture, they fail to accurately capture the economics of these newer platforms~\cite{burckhardt2021formale,lin2021formale}.

%% file: sections/9_conclusion.tex
\section{Conclusion}

As serverless computing continues to grow in importance for both industry and academia, understanding the underlying platform architectures is highly relevant.
In this paper, we investigated the architectural paradigms of commercial FaaS platforms and their impact on performance.
We provided a comprehensive analysis of seven major providers -- AWS Lambda, Cloudflare Workers, Deno Deploy, Fastly Edge Compute, Fly.io, Google Cloud Run Functions, and Oracle Cloud Infrastructure Functions -- by synthesizing extensive publicly available data to yield detailed insights into their system designs.
To bridge the gap between architectural theory and real-world execution, we developed a benchmarking suite and ran it from November 2025 to January 2026, encompassing \SI{38165688} function calls. 
Our empirical evaluation reveals a clear differentiation between first-generation platforms, which rely on containerized, centralized execution, and a new second generation.
By leveraging lightweight isolates and edge deployments, these newer platforms pivot away from providing highly configurable, resource-rich environments.
Instead, they focus on executing short, lightweight functions at a global scale.

Our findings demonstrate that this second generation does not simply replace or universally outperform the first; rather, it targets different use cases.
While first-generation FaaS remains a highly capable and necessary choice for flexible, resource-intensive, and complex workloads, the second generation excels in distributed, low-latency scenarios. 
For these specific edge-native workloads, the performance improvements are substantial: warm request latencies drop from \SI{\sim 40}{ms} to \SI{\sim 10}{ms}. 
Furthermore, cold starts—once a defining bottleneck in serverless application development—cease to be a key consideration, becoming merely an afterthought. 
Although the restrictive execution environments of second-generation platforms limit programming language choices and available APIs, their architectures effectively solve long-standing challenges in serverless computing, such as the double billing problem. 
To encourage future research and further exploration into the diverging paths of FaaS architectures, we release our benchmarking suite and the complete dataset as open source.